\documentclass{elsart3}
\journal{Physica B}
\usepackage{graphicx}
\usepackage{amssymb}
\begin{document}

\begin{frontmatter}

% Title, authors and addresses

% use the thanksref command within \title, \author or \address for footnotes;
% use the corauthref command within \author for corresponding author footnotes;
% use the ead command for the email address,
% and the form \ead[url] for the home page:
% \title{Title\thanksref{label1}}
% \thanks[label1]{}
% \author{Name\corauthref{cor1}\thanksref{label2}}
% \ead{email address}
% \ead[url]{home page}
% \thanks[label2]{}
% \corauth[cor1]{}
% \address{Address\thanksref{label3}}
% \thanks[label3]{}

\title{Heating process in the pre-Breakdown regime of the Quantum Hall Effect : a size dependent effect}

% use optional labels to link authors explicitly to addresses:
% \author[label1,label2]{}
% \address[label1]{}
% \address[label2]{}
\author{Y.M. Meziani}%~\corauthref{cor}, }
\ead{yahya@ges.univ-montp2.fr}
\author{, C. Chaubet, B. Jouault, S. Bonifacie and A. Raymond}
%\corauth[cor]{Mail : yahya@ges.univ-montp2.fr}
\address{Universit\'e Montpellier 2, Groupe d'Etude des Semiconducteurs, Place Eug\`ene Bataillon, 34096 Montpellier, France}
\author{W. Poirier and F. Piquemal}

\address{Bureau National de M\'etrologie, Laboratoire National d'Essais
(BNM-LNE), 33 avenue du g\'en\'eral Leclerc 92260 Fontenay aux Roses, France}
\begin{abstract}
Our study presents experimental measurements of the contact and longitudinal voltage drops in Hall bars, as a function of the current amplitude. We are interested in the heating phenomenon which takes place before the breakdown of the quantum Hall effect, i.e. the pre-breakdown regime. Two types of samples has been investigated, at low temperature (4.2 and 1.5K) and high magnetic field (up to 13 T). The Hall bars have several different widths, and our observations clearly demonstrate that the size of the sample influences the heating phenomenon. By measuring the critical currents of both contact and longitudinal voltages, as a function of the filling factor (around $i=2$), we highlight the presence of a high electric field domain near the source contact, which is observable only in samples whose width is smaller than 400 microns. % Text of abstract
\end{abstract}

\end{frontmatter}

\section{Introduction}
\label{Intro}
\par Ten years after its discovery by Klaus von Klitzing~\cite{Klitzing80}, the Quantum Hall Effect (QHE) was used to conserve the unit of
the electric resistance. In the plateau regime, $\rho_{xx}=0$ and $\rho_{xy} = R_K/i$ where $R_K=h/e^2$ (h - Planck constant, e - electron charge, $i$ - filling factor (number of occupied Landau levels)) for the temperature approaching zero. The value of the von Klitzing constant $R_K$ has been used as the standard of resistance by many laboratory of metrology, because the relative precision of its measurement is better than $10^{-9}$. However, the current must be very well controled because a sudden breakdown of the QHE occurs if it exceeds a certain limit. Above this critical current, the system becomes unstable, and the longitudinal resistance $\rho_{xx}$ changes abrubtly by several orders of magnitude~\cite{Ebert83}. This effect has been attracting interest by a steadily community of researchers~\cite{Nachtwei98,Chaubet98,Tsemekhman97,Eaves01}. In the present review we are interested in the pre-breakdown regime, i.e. in the heating phenomenon 
which is observable before breakdown of QHE. We studied precisely for the first time, the voltage drop between the source contact and the adjacent voltage contact in the QHE regime, using a well specified configuration for the measurement. This paper is organized as follows. After this introduction, we present the samples and the set up of our measurements. In section 3, we present and discuss the results. We show that the dimensions of the samples influence drastically the measurements of the contact voltage drop in this pre-breakdown regime. 
%==============================================
\section{Samples and Configurations}
\subsection{Samples}
\label{Samples}
%==============================================
Samples investigated in this review are GaAs/Ga\-AlAs heterojunctions grown on 3 inches wafers by Metal Organic Chemical Vapor Deposition
(MOCVD) technique. This procedure allows us to obtain a good homogeneity of the electronic density. Starting from the substrate, a 600 nm thick undoped GaAs buffer layer is firstly deposited. It is followed by an undoped Al$_{0.28}$Ga$_{0.72}$As spacer layer whose thickness is respectively 22 nm and 14.5 nm for PL175 and PL173 heterostructures. Then a 40 nm thick 10$^{18}$ cm$^{-3}$ Si-doped Al$_x$Ga$_{1-x}$As layer is realized, with a gradual decrease of x up to 0.28 for PL175 and a homogenous value (x=0.28) for PL173~\cite{Piquemal99,Poirier02,Piquemal93}. Finally an n-type 12 nm GaAs cap layer covers the structure to improve the quality of ohmic contacts. After the realization of the 300 nm thick delimiting mesa, the AuGeNi ohmic contacts are evaporated and then annealed at 450 $^\mathrm{o}$C.
\par
\begin{figure}[htbp]
\begin{center}
\includegraphics[scale=.40]{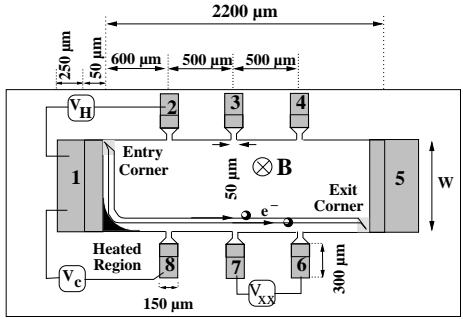}
\caption{Geometry of Hall bar. W is the width of the Hall bar. Two current contacts
are used as a drain and a source (contacts 1 and 5). Six other
lateral contact are used for measuring voltage drops. The magnetic polarity is positive here.}
\label{fig:hallbar}
\end{center}
\end{figure}

 All samples were processed into a Hall bar. They have six independent lateral contacts in addition to the source and drain contacts, as described in Fig.~\ref{fig:hallbar}. Table~\ref{tab:samples} resume the characteristics of the samples. They were mounted on a sample holder and placed inside a Variable Temperature Insert (VTI), in a 16 T superconducting magnet. All experiments were performed at 4.2 K and 1.5 K.

\begin{table}[htbp]
\begin{center}
\caption{Samples Characteristics (i is the filling factor).}
\label{tab:samples}
\begin{tabular}{||c c c c c||}
\hline
\hline
Wafer & $\mathrm{N_s}$ & $\mathrm{\mu}$ & B(i=2) & Width W\\
& 10$\mathrm{^{15}}$ m$\mathrm{^{-2}}$ & (m$\mathrm{^2}$/V.s) & (T) &
($\mathrm{\mu}$m)\\
\hline
PL173 & 3.3 & 50 & 6.8 & 200, 400 and 800\\
\hline
PL175 & 4.3 & 42.5 & 9 & 200, 400 and 1600\\
\hline
\hline
\end{tabular}
%\caption{Samples Characteristics (i is the filling factor).}
%\label{tab:samples}
\end{center}
\end{table}
%=====================================================
\subsection{Configuration for the measurements}
%=====================================================
All the measurements have been carried out only at the i=2 plateau. In the QHE regime, the magnetic field $B$ strongly bends the potential profile. If a D.C current is applied between contacts {\bf1} (source) and {\bf 5} (drain) and for B positive (+B, see Fig. 1), electrons enter the two dimensional electron gas (2DEG) by one corner of the source (Entry Corner) and leave it by the opposite corner (Exit Corner) of the drain. Around the center of the plateau $i=2$, one measures the Hall voltage between the current contacts {\bf 1} and {\bf 2} : V$_H$=V$_{12}$=(h/2e$^2$) and V$_c$=0 for low current levels. A bref illustration is presented in Fig.~\ref{fig:hallbar}.
\par The existence of electron entry and exit corner has been reported for the first time by Klaus et al.~\cite{Klaus91} using an optical FIR technique that allows to observe the high electric field zones in the sample. They have shown that the Hall potential is concentrated on a very narrow region whose caracteristic length is 10 $\mathrm{\mu}$m. Kawano et al.~\cite{Kawano00} used a spatially resolved detector to measure the FIR emission of the sample under QHE regime, and reported on the same location of hot spots. However at higher current levels, a new region, located in the vicinity of the source contact, emits also a FIR radiation. This additional signal dominates at high level currents. This shows that a heating process takes place along the edge source contat, above a critical current. This specific heating process exists while the sample is still in the QHE regime and constitutes the subject of our investigations.\\

Starting from the configuration of measurement shown in Fig.1, we have used the four different configurations that can be deduced from this one by reversing the polarities of both current and magnetic field. In the Buttiker Landauer notations, one can easily name the four configurations by using the contacts labels shown in Fig.~\ref{fig:hallbar} : V$_{ij,kl}$ is the voltage drop V$_{kl}$ measured when the current I$_{ij}$ is injected from i to j. The contact resistance is therefore measured in the four configurations : V$_{51,18}$(B), V$_{15,54}$(B), V$_{51,12}$(-B) and V$_{15,56}$(-B). The corresponding longitudinal voltage drops are : V$_{51,76}$(B), V$_{15,32}$(B), V$_{51,34}$(-B) and V$_{15,78}$(-B). In our experiments, we have studied the four configurations for each Hall bars, and obtained well reproducible features of these voltage drops. We can then deduce that the pre-breakdown regime exhibits very stable characteristics and is therefore the consequence of a specific heating regime that we will discuss now.
\par Measurement of V$_c$ and V$_{xx}$ were made simultaneously as a function of the current and for fixed magnetic field, so that we can compare the critical current for the threshold of V$_c$ and V$_{xx}$ for different Hall bar widths . For a given value of magnetic field, which corresponds approximatively to the center of plateau i=2, we did measurements of V$_c$(I) and V$_{xx}$(I). Then we have slightly changed the value of B and make again the same measure again, always taking care to never overpass the breakdown, that would have induced hysterisis and non reproductible data as yet observed~\cite{Cage90,Boella94,Ahlers90}.
%================================
\section{Results and Discussion}
\label{Results}
%===============================
\subsection{Medium size samples}
\par
\begin{figure}[htbp]
\begin{center}
\includegraphics*[scale=.2]{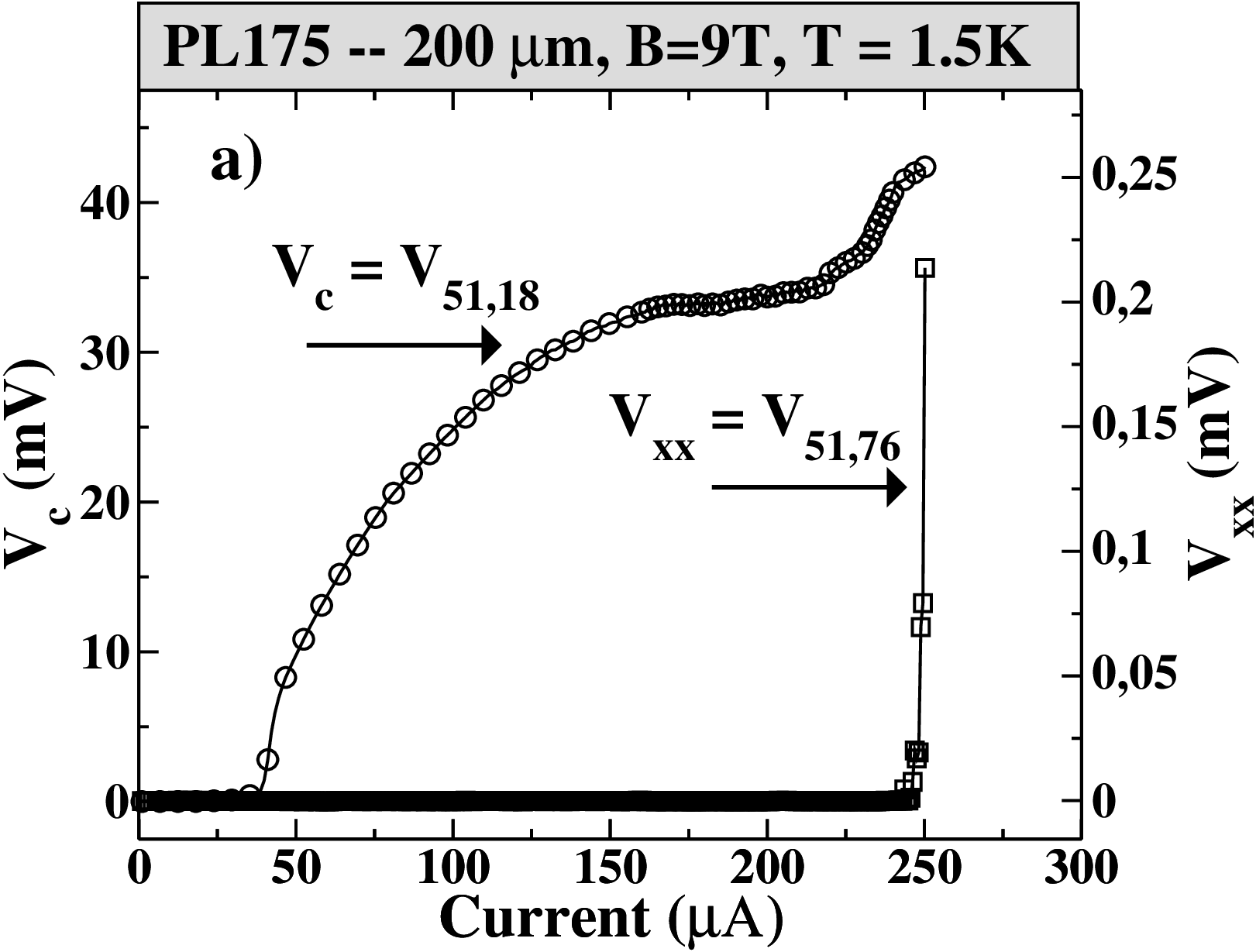}
\includegraphics*[scale=.2]{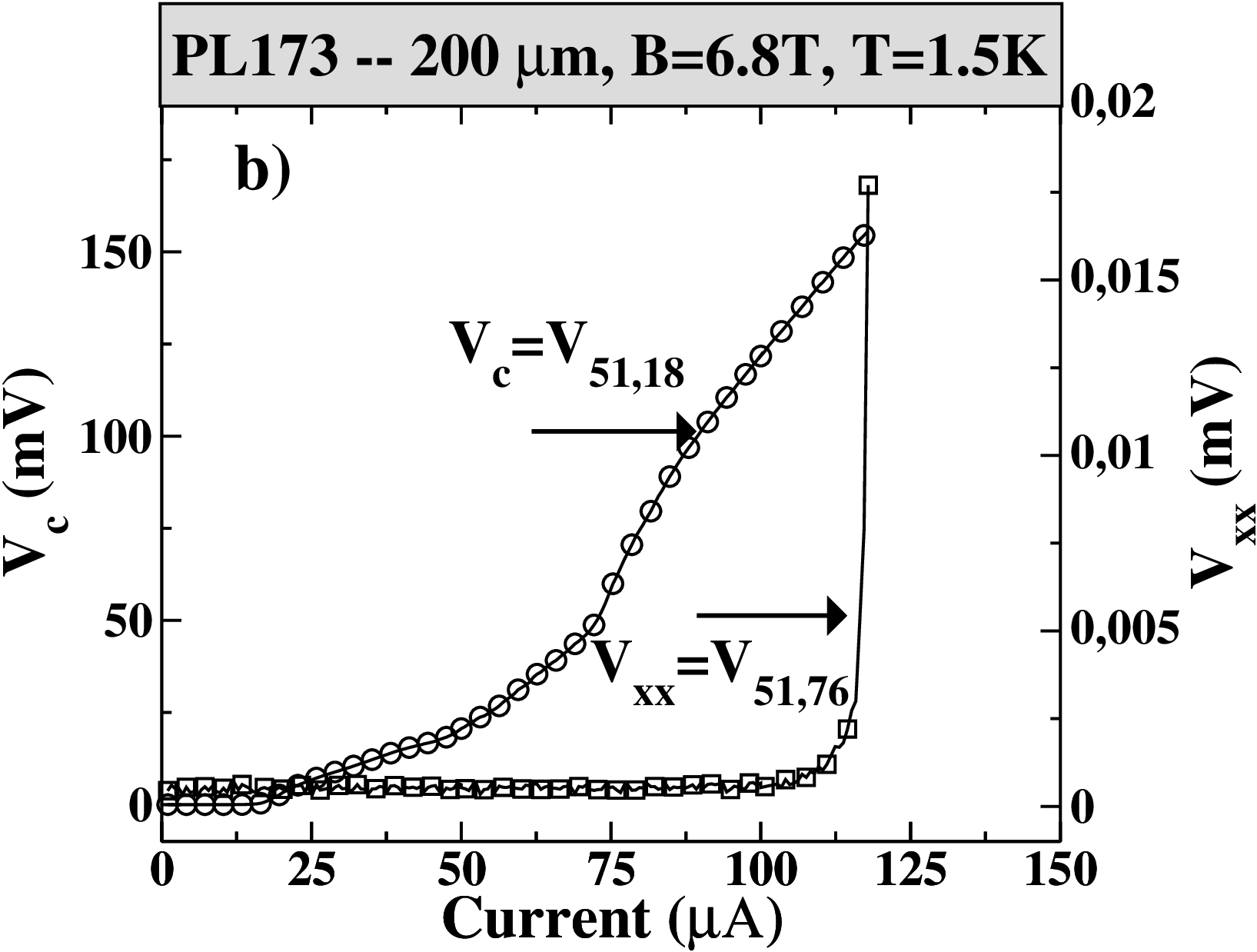}
\caption{Measurement of $\mathrm{V_c=V_{51,18}}$ and
$\mathrm{V_{xx}=V_{51,76}}$. {\bf a)} Sample PL175--200 $\mu$m at B=9T and
T=1.5K. {\bf b)} Sample PL173--200 $\mu$m at B=6.8T and T=1.5K. For both cases, critical current of V$_c$ is mush lower than those of V$_{xx}$.}
\label{fig:VcVxx}
\end{center}
\end{figure}

 We first comment the results obtained for the samples of width 200 $\mu$m and 400 $\mu$m. In Fig.~\ref{fig:VcVxx} we have reported the measurements of $\mathrm{V_c=V_{51,18}}$ and $\mathrm{V_{xx}=V_{51,76}}$ versus current, for samples PL175 and PL173 of width W=200 $\mu$m. The temperature was T=1.5K. First, we observe an abrupt jump of voltage drop $V_c$ for both samples that allows us to define a critical current I$_c$ for V$_c$ : I$_c$=40 $\mu$A for PL175 and I$_c$=25 $\mu$A for PL173. This shows that a local breakdown occurs at the vicinity of the current contact, well before the onset of V$_{xx}$ in the entire 2DEG. Indeed the critical current for the breakdown (called I$_b$) is I$_b$=250 $\mu$A for PL175 and 120 $\mu$A for PL173. Kawano et al have already commented the particular behavior of this contact resistance, that we clarify here by a systematic study. In fact, in their study, Kawano et al. have precisely reported on the Far Infra Red emission which appears in the same zone and for the same current level. It is therefore clear that a heating phenomenon due to the presence of high electric field takes place near the source, in the pre-breakdown regime. 

\par We will discuss in the following the size of this heating region. It is necessarily an extended area. Indeed, it is established that the breakdown needs to developp a sufficiently large domain of high electric field. In that case, electron holes pairs can be created according to the Boot-strap electron heating model~\cite{Komiyama96,Kawaguchi95}. By investigating smaller samples (width smaller than 100$\mu$m) Komiyama et al. have deduced that the minimum length scale to developp the breakdown is larger than 100$\mu$m~\cite{Komiyama96}. 

\par
\begin{figure}[htbp]
\begin{center}
\includegraphics*[scale=.2]{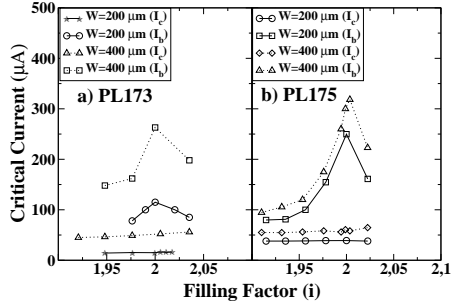}
\caption{Variation of critical current of V$_c$ and V$_{xx}$ as a function of filling factor for samples PL173 and PL175 with diffrents width. One observes that the critical current of V$_c$ is independante of filling factor in the plateau regime.}
\label{fig:Icnu}
\end{center}
\end{figure}

 We think that the high field domain reaches the edge of the sample because the resistance between the current contact and the first voltage contact is affected. To prove this more convincingly, we have dressed the graph of the variation of the critical current for both contact voltage drop and longitudinal voltage drop versus filling factor ($i$). Samples PL175 and PL173 with differents width Hall bar are concerned (see Fig.~\ref{fig:Icnu}). One observes, for all samples, that the critical current of longitudinal voltage drop (I$_b$) is maximum at the center of plateau ($i=2$). Nachtwei et al.~\cite{Nachtwei97,Liu98} have also observed same results in 2D systems with periodic and aperiodic antidots array. Here, for the first time, we observe that the critical current of the contact resistance (I$_c$) does not depend on the magnetic field (filling factor) in the plateau regime. Then it is clear that the local breakdown near the contact is different from the breakdown in the entire 2DEG, due to the presence of a domain of high electric field. 

\subsection{Large samples}

\par
\begin{figure}[htbp]
\begin{center}
\includegraphics*[scale=.2]{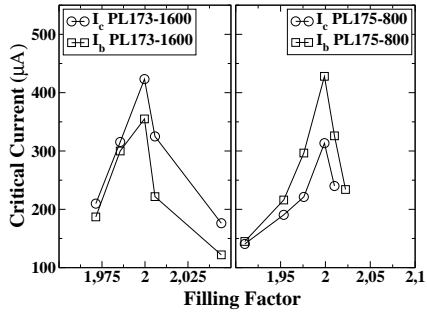}
\caption{Same of figure~\ref{fig:Icnu} but in this case samples have alarger
width : {\bf a)} Sample PL173 with W=1600 $\mu$m. {\bf b)} Sample
PL175 with W=800 $\mu$m.}
\label{fig:IcIbnu}
\end{center}
\end{figure}

 A particular behavior for I$_c$(i) and I$_b$(i) was observed for wide Hall bars. In Figure~\ref{fig:IcIbnu}, we report behavior of criticals currents I$_c$ and I$_b$ versus filling factor $i$. For sample PL175 with W=800 $\mu$m, we observe that I$_c$ and I$_b$ are maximum in center of the plateau. For sample PL173 with W=1600 $\mu$m, we have the same result but the value of I$_c$ is now higher than that of I$_b$.
\par It is obvious then that the critical currents $I_c$ and $I_b$ have the same variation with filling factor. This proves that the heating process in both regions are identical. Hence the high field zone has a size which is smaller than 800$\mu$m, because it does not affect the equilibrium of edge states on the edge of the samples. Then we can have an estimated length for this high electric zone, which is certainly several hundreds of microns, but less than 800 $\mu$m : tipycally around 500$\mu$m. This constitutes a new characteristic dimension of the heated region in the pre-breakdown regime.

\section{Conclusion}
\label{Conclusion}
A new measurement protocol has been used to characterize the pre-breakdown regime of the QHE. We have characterized for the first time the voltage drop across the contact and dressed the graph of the variation for its critical current with the filling factor. Large samples exhibited a different behavior than medium size ones. We can therefore deduce an estimated length for the extended domain of high electric field, which is appears near the source. The propagation and extension of the high electric field domain remains of course to be further studied. We propose that the Bootstrap electron heating model should be proper to describe this phenomenon. This work was granted by the Bureau National de M\'etrologie.
% main text

% The Appendices part is started with the command \appendix;
% appendix sections are then done as normal sections
% \appendix

% \section{}
% \label{}


\begin{thebibliography}{00}
\bibitem{Klitzing80} K. von Klittzing, G. Dorda, M. Pepper,
Phys. Rev. Lett. 34 (1980) 494.
\bibitem{Ebert83} G. Ebert, K. von Klittzing, K. Ploog, G. Weimann,
J. Phys. C. 16 (1983) 5441.
\bibitem{Nachtwei98} G. Nachtwei, Physica E 4 (1999) 79-101.
\bibitem{Chaubet98} C. Chaubet and F. Geniet, Phys. Rev. B, 58 (1998) 13015.
\bibitem{Tsemekhman97} V. Tsemekhman, K. Tsemekhman, C. Wexler,
J. H. Han and D. J. Thouless, Phys. Rev. B 55 (1997) R10201.
\bibitem{Eaves01} L. Eaves, Physica B, 298 (2001) 1-7.
\bibitem{Piquemal99} F. Piquemal, Bulletin du BNM, 116, 2 (1999).
\bibitem{Poirier02} W. Poirier, A. Bounouh, K. Haayashi, H. Fhima,
F. Piquemal, G. Genev\`es and J. P. Andr\`e, J. Appl. Phys. 92, 5
(2002).
\bibitem{Piquemal93} F. Piquemal, G. Genev\`es, F. Delahaye,
J. P. Andr\`e, J. N. Patillon and P. Frijlink, IEEE
Trans. Instrum. Meas. 42 (1993) 264-268.
\bibitem{Meziani02} Y.M. Meziani, C. Chaubet, S. Bonifacie,
B. Jouault, A. Raymond, W. Poirier and F. Piquemal,
Proc. Int. Conf. on High Magnetic Fields in Semiconductors Physics,
Oxford, U.K 2002.
\bibitem{Klaus91} U. Klaus, W. Dietsche, K. von Klitzing a,d K. Ploog,
Z. Phys. B 82 (1991) 351.
\bibitem{Kawano00} Y. Kawano and S. Komiyama, Phys. Rev. B 61 (2000)
2961.
\bibitem{Cage90} M. E. Cage, G. Marullo Reedtz, Y. D. Yu and C. T. van
Degrift, Semicond. Sci. Technol. 5 (1990) 351.
\bibitem{Boella94} G. Boella, L. Cordiali, G. Marullo Reedtz,
D. Allasia, GM. Truccato and C. Villavecchia, Phys. Rev. B 50 (1994)
7608.
\bibitem{Ahlers90} F. J. Ahlers, G. Hein, H. Scherer, L. Bliek,
H. Nickel, R. Losch and W. Schlapp, Semicond. Sci. Technol. 8 (1990)
2062.
\bibitem{Nachtwei97} G. Nachtwei, G. L\"{u}tjering, D. Weiss, Z.H. Liu, K. von
Klitzing, C.T. Foxon, Phys. Rev. B 55 (1997) 6731.
\bibitem{Liu98} G. Nachtwei, Z.H. Liu, G. L\"{u}tjering, R.R. Gerhards,
D. Weiss, , K. von Klitzing, K. Eberl, Phys. Rev. B 57 (1998) 9937.
\bibitem{Komiyama96} S. Komiyama, Y. Kawaguchi, T. Osada and Y. Shiraki,
Phys. Rev. Lett. 77 (1996) 558.
\bibitem{Kawaguchi95}Y. Kawaguchi, F. Hayashi, S. Komiyama, T. Osada,
Y. Shiraki and R. Itoh, Jpn J. Appl. Phys. 34 (1995) 7608.
% Text of bibliographic item

% notes:
% \bibitem{label} \note

% subbibitems:
% \begin{subbibitems}{label}
% \bibitem{label1}
% \bibitem{label2}
% If there is a note, it should come last:
% \bibitem{label3} \note
% \end{subbibitems}

%\bibitem{}

\end{thebibliography}
\end{document}